\documentclass[epj,twocolumns]{svjour_mod}

\usepackage{latexsym}
\usepackage{graphics}
\usepackage{amssymb}
\usepackage[colorlinks,citecolor=blue,urlcolor=blue,linkcolor=blue]{hyperref}
\usepackage{hyphenat}
\usepackage{amsmath}
\usepackage{cite}
\usepackage{relsize}
\usepackage{trimclip}
\usepackage{booktabs} 
\usepackage{tabularx}
\usepackage{multirow}
\usepackage{fixmath}
\usepackage{tikz}
\usetikzlibrary{calc}
\usepackage{placeins}

\newcommand{\ptcut}{p_{T,{\rm cut}}^h}
\newcolumntype{U}{>{\centering\arraybackslash}X}
\makeatletter
\g@addto@macro\bfseries{\boldmath}
\makeatother

\begin{document}

%Title
\title{Charged hadron fragmentation functions from collider data}

%Authors
\author{The NNPDF Collaboration: V.~Bertone \inst{1,2} 
\and N.P.~Hartland\inst{1,2}
\and E.R.~Nocera\inst{3}
\and J.~Rojo\inst{1,2}
\and L.~Rottoli\inst{4}}      

%Affiliations
\institute{Department of Physics and Astronomy, Vrije Universiteit
  Amsterdam, 1081 HV
  Amsterdam, The Netherlands.
\and Nikhef Theory Group, Science Park 105, 1098 XG Amsterdam, The
  Netherlands.
\and Higgs Centre for Theoretical Physics, School of Physics and
  Astronomy, University of Edinburgh, EH9 3FD, United Kingdom.
\and Rudolf Peierls Centre for Theoretical Physics, University of Oxford, 
  Clarendon Laboratory, Parks Road, Oxford OX1 3PU, United Kingdom.}

%Date
\date{}

%Abstract
\abstract{We present NNFF1.1h, a new determination of unidentified
  charged\hyp{}hadron fragmentation functions (FFs) and their
  uncertainties.  Experimental measurements of
  transverse\hyp{}momentum distributions for charged\hyp{}hadron
  production in proton\hyp{}(anti)proton collisions at the Tevatron
  and at the LHC are used to constrain a set of FFs originally
  determined from electron\hyp{}positron annihilation data. Our
  analysis is performed at next\hyp{}to\hyp{}leading order in
  perturbative quantum chromodynamics. We find that the
  hadron\hyp{}collider data is consistent with the
  electron\hyp{}positron data and that it significantly constrains the
  gluon FF. We verify the reliability of our results upon our choice
  of the kinematic cut in the hadron transverse momentum applied to
  the hadron\hyp{}collider data and their consistency with NNFF1.0,
  our previous determination of the FFs of charged pions, kaons, and
  protons/antiprotons.  }

%PACS

%Frontpage
\maketitle

%Body of the paper
\section{Introduction}
\label{sec:intro}

The determination of the collinear unpolarised fragmentation functions
(FFs) of neutral and charged hadrons has been a topic of active research 
in the last decade~\cite{Metz:2016swz}.
FFs describe how coloured partons are turned into hadrons and can be
regarded as the final\hyp{}state counterparts of the parton
distribution functions (PDFs)~\cite{Gao:2017yyd}.
Since FFs are non\hyp{}perturbative quantities in quantum
chromodynamics (QCD), they need to be determined from an analysis of
experimental data.

The recent interest in FFs stems from the copious amount of precise
measurements that have been and are currently being collected for
different processes at various centre\hyp{}of\hyp{}mass energies.
These include data for hadron production in: single\hyp{}inclusive
$e^+e^-$ annihilation (SIA) (recently measured by
BELLE~\cite{Leitgab:2013qh,Seidl:2015lla} and
BABAR~\cite{Lees:2013rqd}), semi\hyp{}inclusive deep\hyp{}inelastic
scattering (SIDIS) (recently measured by
HERMES~\cite{Airapetian:2012ki} and
COMPASS~\cite{Adolph:2016bga,Adolph:2016bwc}) and
proton\hyp{}(anti)proton ($pp$) collisions (measured, {\it e.g.}, by
CDF~\cite{Abe:1988yu,Aaltonen:2009ne} at the Tevatron,
STAR~\cite{Adamczyk:2013yvv} and PHENIX~\cite{Adare:2007dg} at RHIC
and CMS~\cite{Chatrchyan:2011av,CMS:2012aa} and
ALICE~\cite{Abelev:2013ala} at the LHC).
These measurements span a wide range in energy and momentum fraction
and are sensitive to different partonic combinations.
Therefore, they offer a unique opportunity to determine FFs with an
unprecedented accuracy.

Several analyses exploited some of these measurements to constrain the FFs of 
the lightest charged hadrons, {\it i.e.} $\pi^\pm$, $K^\pm$, and 
$p/\overline{p}$.
Among the most recent studies, the HKKS16~\cite{Hirai:2016loo}, 
JAM16~\cite{Sato:2016wqj}, and NNFF1.0~\cite{Bertone:2017tyb} analyses 
are based on SIA data only.
A global determination of the charged pion and kaon FFs was carried
out in Refs.~\cite{deFlorian:2014xna,deFlorian:2017lwf}, where SIDIS
and $pp$ data was also included.
The FFs of heavier hadrons, such as
$D^*$~\cite{Anderle:2017cgl,Soleymaninia:2017xhc},
$\Lambda$~\cite{Albino:2005mv,Albino:2008fy} and
$\eta$~\cite{Aidala:2010bn}, were also studied, mostly from SIA measurements,
although available data is in general scarcer than for light hadrons.

A further family of FFs with phenomenological relevance are those of the
unidentified charged hadrons.
They can be regarded as the sum of the FFs of all charged hadrons that
can be produced in the fragmentation of a given parton.
These FFs find application, for example, in the description of the
charged\hyp{}particle spectra measured in proton\hyp{}ion and
ion\hyp{}ion collisions, which are actively investigated by current
RHIC~\cite{Adams:2005dq} and LHC~\cite{Abreu:2007kv} heavy\hyp{}ion programs.

Despite the fair amount of measurements sensitive to unidentified
charged\hyp{}hadron FFs, they have received less attention as compared
to identified charged\hyp{}hadron FFs.
As a matter of fact,  only a few extractions have been
carried out until recently~\cite{Kretzer:2000yf,
  Kniehl:2000fe,Bourhis:2000gs,deFlorian:2007ekg}. The analysis of
Ref.~\cite{deFlorian:2007ekg} is the only fit based on SIA, SIDIS and
$pp$ data, while all others are based on SIA data only.
These FF sets were extracted some time ago from older measurements and it has
been shown~\cite{dEnterria:2013sgr} that they do not describe the more recent
transverse\hyp{}momentum charged\hyp{}particle spectra measured at the Tevatron
and the LHC\@.

New analyses of unidentified charged\hyp{}hadron FFs have been presented
recently~\cite{Nocera:2017gbk,Soleymaninia:2018uiv} based only upon SIA data.
In particular, the determination in Ref.~\cite{Nocera:2017gbk} was performed
using the NNPDF fitting
methodology~\cite{DelDebbio:2007ee,Ball:2008by,Ball:2010de} designed to provide
a statistically sound representation of experimental uncertainties with minimal
procedural bias.
As the SIA dataset used in this analysis has little power to constrain the
gluon FF, the resulting gluon distribution was found to be affected by large
uncertainties, within which the discrepancy in the description of $pp$ data
reported in Ref.~\cite{dEnterria:2013sgr} could be mitigated.

The purpose of this paper is to complement the analysis of
Ref.~\cite{Nocera:2017gbk} with the most recent measurements of the
transverse\hyp{}momentum charged\hyp{}hadron spectra in $pp$
collisions.
These measurements are directly sensitive to the so far
poorly known gluon fragmentation, therefore their inclusion in a
fit is expected to provide a stringent constraint on this
distribution.
The $pp$ data is included by means of Bayesian
reweighting~\cite{Ball:2010gb,Ball:2011gg,zahari_kassabov_2019_2571601}.
The result, NNFF1.1h, is a new set of FFs for unidentified charged
hadrons from a global analysis of SIA and $pp$ data.

The paper is organised as follows.
In Sect.~\ref{sec:input}, we present the data set included in this
analysis and discuss how the theoretical predictions of the
corresponding observables are computed.
In Sect.~\ref{sec:results}, we present the main results of our analysis.
Specifically, in Sect.~\ref{sec:impact} we discuss the quality of the
fit and the impact of the hadron\hyp{}collider data on the FFs; in
Sect.~\ref{sec:pertacc} we motivate our choice for the kinematic cut
on the transverse\hyp{}momentum of the final\hyp{}state hadron applied
to $pp$ data; and in Sect.~\ref{sec:consistency} we assess the
consistency of the current determination with
NNFF1.0~\cite{Bertone:2017tyb}, our previous analysis of FFs for
charged pions, kaons and protons/antiprotons.
A summary and an outlook are given in Sect.~\ref{sec:summary}.

\section{Experimental and theoretical input}\label{sec:input}

In this section, we present the SIA and $pp$ data sets used in this work and
discuss the theoretical calculation of the corresponding observables.

\subsection{The data set}\label{sec:dataset}

In this analysis we include all available SIA measurements from LEP
(ALEPH~\cite{Buskulic:1995aw}, DELPHI~\cite{Abreu:1998vq,Abreu:1997ir} and
OPAL~\cite{Ackerstaff:1998hz,Akers:1995wt}), PETRA
(TASSO~\cite{Braunschweig:1990yd}), PEP (TPC~\cite{Aihara:1988su}) and
SLC (SLD~\cite{Abe:2003iy}).
These measurements consist of cross sections differential in the
scaling variable $z=2(p^h\cdot q)/Q^2$, where $p^h$ is the
four\hyp{}momentum of the final\hyp{}state hadron, $q$ is the
four\hyp{}momentum of the exchanged virtual gauge boson and
$Q\equiv\sqrt{q^2}$.
They are normalised to the total cross section for inclusive
electron\hyp{}positron annihilation into hadrons, $\sigma_{\rm tot}$.
Besides measurements based on inclusive samples, which contain all
quark flavours, we also include measurements based on
flavour\hyp{}enriched (or tagged) $uds$-, $c$- and $b$-quark samples
from DELPHI~\cite{Abreu:1998vq,Abreu:1997ir},
OPAL~\cite{Ackerstaff:1998hz} and SLD~\cite{Abe:2003iy}.
This data set is then equivalent to that of the identified charged pions, kaons
and protons/antiprotons set used in NNFF1.0.
We refer the reader to Ref.~\cite{Bertone:2017tyb} for a detailed
discussion.

In contrast with identified light hadrons, separate measurements of
the longitudinal contribution to the differential cross sections are
available for unidentified charged hadrons.
We include both inclusive measurements, provided by
DELPHI~\cite{Abreu:1997ir} and OPAL~\cite{Akers:1995wt}, and $uds$-
and $b$\hyp{}tagged measurements, provided by
DELPHI~\cite{Abreu:1997ir}.

The features of the SIA measurements included in this analysis, such
as the centre\hyp{}of\hyp{}mass energy $\sqrt{s}$, the number of data points
for each experiment and their references, are summarised in Table~1 of
Ref.~\cite{Nocera:2017gbk}.
Our SIA data set mostly overlaps that of previous
analyses~\cite{Kretzer:2000yf,Kniehl:2000fe,Bourhis:2000gs,deFlorian:2007ekg,
  Nocera:2017gbk,Soleymaninia:2018uiv}.

Concerning $pp$ data, we include all available measurements from the
Tevatron (CDF~\cite{Abe:1988yu,Aaltonen:2009ne}) and the LHC
(ALICE~\cite{Abelev:2013ala} and CMS~\cite{Chatrchyan:2011av,CMS:2012aa}).
They consist of cross sections differential in the momentum of the
final\hyp{}state hadron, $p^h$, presented as a function of its transverse
component $p_T^h$ at different centre\hyp{}of\hyp{}mass energies $\sqrt{s}$.
Specifically, we include CDF data at 1.80 TeV~\cite{Abe:1988yu} and
1.96 TeV~\cite{Aaltonen:2009ne}, CMS data at 0.9
TeV~\cite{Chatrchyan:2011av}, 2.76 TeV~\cite{CMS:2012aa} and 7
TeV~\cite{Chatrchyan:2011av}, and ALICE data at 0.9 TeV, 2.76 TeV, and
7 TeV~\cite{Abelev:2013ala}.
The covered rapidity range is $|\eta|<1$ for CDF and CMS and
$|\eta|<0.8$ for ALICE\@.
The CMS and ALICE data is used here for the first time to constrain FFs.

We do not consider older measurements performed by the
UA1~\cite{Albajar:1989an,Arnison:1982ed,Bocquet:1995jr} and
UA2~\cite{Banner:1984wh} experiments at the Sp$\overline{\rm p}$S nor
those by the PHENIX experiment~\cite{Adler:2005in} at RHIC\@.
These measurements mostly cover the low-$p_T^h$ region, where large
missing higher\hyp{}order corrections affect the theoretical
predictions.
They would therefore be almost completely excluded by our kinematic
cuts (see Sect.~\ref{sec:theory}).
These measurements were also found to be poorly
described when included in a global fit of FFs~\cite{deFlorian:2007ekg}.

The features of our $pp$ data set are summarised in Table~\ref{tab:results},
where we specify the name of each experiment, the publication reference, the
centre\hyp{}of\hyp{}mass energy $\sqrt{s}$ and the number of data points,
$N_{\rm dat}$.

\subsection{Theoretical calculations}\label{sec:theory}

The normalised SIA total (longitudinal) cross section can be expressed
in a factorised form as
\begin{equation}\label{eq:SIAxsec}
  \frac{1}{\sigma_{\rm tot}}\frac{d\sigma^{h^\pm}_{2(L)}}{dz}(z,Q)
  =
  \frac{4\pi\alpha^2}{\sigma_{\rm tot} Q^2}\sum_{l} C_{2(L)}^l(z,Q)\otimes D_l^{h^\pm}(z,Q)\,,
\end{equation}
where $h^\pm$ denotes the sum of unidentified charged hadrons,
$h^\pm=h^++h^-$, $\alpha$ is the quantum electrodynamics (QED)
coupling constant and $\otimes$ represents the convolution pro\-duct
between the perturbative total (longitudinal) coefficient functions
$C_{2(L)}^l$ and the non\hyp{}perturbative FFs $D_l^{h^\pm}$
associated to the parton $l$.
The sum over $l$ in Eq.~\eqref{eq:SIAxsec} runs over all active
partons at the scale $Q$.

As discussed in Sect.~3.1 of Ref.~\cite{Bertone:2017tyb}, the
observable defined in Eq.~\eqref{eq:SIAxsec} is sensitive only to a
limited number of quark FF combinations and to the gluon FF\@.
In the case of the quark FFs, SIA measurements provide limited sensitivity to
the separation between the different light-quark FFs, while a direct handle on
the separation between light- and heavy-quark FFs is provided by the
flavour-tagged data.
The gluon FF is poorly constrained by the total SIA cross
sections $d\sigma^{h^\pm}_{2}/dz$.
The reason being that the total coefficient function of the gluon,
$C_{2}^g$, receives its leading\hyp{}order (LO) contribution at
$\mathcal{O}(\alpha_s)$, while that of the quark, $C_{2}^q$, at
$\mathcal{O}(1)$~\cite{Rijken:1996vr,
  Rijken:1996ns,Mitov:2006wy,Blumlein:2006rr}.
Conversely, the longitudinal cross section $d\sigma^{h^\pm}_{L}/dz$
has a comparable sensitivity to gluon and quark FFs because both
coefficient functions, $C_{L}^g$ and $C_{L}^q$, start at
$\mathcal{O}(\alpha_s)$.
Noticeably, measurements of the longitudinal SIA cross section are
available only for the production of unidentified hadrons.

The numerical computation of the cross sections in Eq.~\eqref{eq:SIAxsec}
and of the evolution of FFs is performed at NLO using {\tt
  APFEL}~\cite{Bertone:2013vaa, Bertone:2015cwa} as in the NNFF1.0 analysis.
In contrast with NNFF1.0, we cannot analyse SIA data at
next\hyp{}to\hyp{}next\hyp{}to\hyp{}leading order (NNLO) as perturbative
corrections to the coefficient functions of the longitudinal cross section in
Eq.~\eqref{eq:SIAxsec} are only known up to $\mathcal{O}(\alpha_s^2)$, {\it
i.e.} NLO.

To avoid regions where small- and large\hyp{}$z$ resummation effects are
sizeable, we impose kinematic cuts on the SIA data.
We adopt the same cuts used in the NNFF1.0 analysis, where data points
below $z_{\rm min}$, with $z_{\rm min}=0.02$ for experiments at
$\sqrt{s}=M_Z$ and $z_{\rm min}=0.075$ for the rest, and above
$z_{\rm max}=0.9$ are excluded from the fit.

Turning to the differential distribution of the final-state hadron in
$pp$ collisions, it can be expressed in a factorised form as
\begin{equation}\label{eq:ppxsec}
  E_h\frac{d^3\sigma^{h^\pm}}{d^3 p^h}
  =
  \sum_{i,j,l}
  K_{ij}^l\otimes
  f_i(x_1,\mu) \otimes f_j(x_2,\mu)
  \otimes D_l^{h^\pm}(z,\mu)\,,
\end{equation}
where $E_h$ and $p^h$ are the energy and the three\hyp{}momentum of
the produced hadron, $f_i(x_1,\mu)$ and $f_j(x_1,\mu)$ are the PDFs of
the colliding hadrons, $D_l^{h^\pm}(z,\mu)$ is the FF of the outgoing
hadron, $K_{ij}^l$ are the perturbative hard cross sections and the
summation runs over all active partons $i,j,k$ at the scale $\mu$.
In principle, the factorisation scale $\mu$ could be chosen
independently for PDFs and FFs, and independently from the
renormalisation scale used in $\alpha_s$.
In practice, our nominal choice is to set all scales equal to the
transverse momentum of the produced hadron, \textit{i.e.} $\mu=p_T^h$.

If heavy\hyp{}quark masses are neglected, as done here, the hard cross
sections $K_{ij}^l$ in Eq.~\eqref{eq:ppxsec} are blind to the quark
flavour of the FF\@.
This implies that the index $l$ distinguishes only whether the
outgoing parton is a gluon or a quark, regardless of its flavour.
This structure can be made explicit by re\hyp{}rewriting
Eq.~\eqref{eq:ppxsec} as
\begin{equation}\label{eq:ppflstruct}
  E_h\frac{d^3\sigma^{h^\pm}}{d^3 p^h}
  =
  \sum_{i,j} f_i \otimes f_j \otimes\left[K_{ij}^g\otimes D_g^{h^\pm}
    + K_{ij}^q\otimes D_\Sigma^{h^\pm}\right]\,,
\end{equation}
where we drop all function dependencies to simplify the notation and define the
singlet FF as
$D_\Sigma^{h^\pm}=\sum_q D_q^{h^\pm}+D_{\bar q}^{h^\pm}$.
The flavour structure of the observable in
Eq.~\eqref{eq:ppflstruct} is therefore such that $pp$ cross-section data is
sensitive only to two independent FF combinations, namely
$D_g^{h^\pm}$ and $D_\Sigma^{h^\pm}$.
This is a subset of the combinations involved in the computation of
the SIA cross sections, see \textit{e.g.} Eq.~(3.1) in
Ref.~\cite{Bertone:2017tyb}.
This property ensures that a prior set of FFs determined from a fit to
SIA data only can be sensibly reweighted with $pp$ cross section data,
as this is not sensitive to any new FF combinations.

The relative contribution of quark and gluon FFs to
Eq.~\eqref{eq:ppxsec} depends on the kinematics.
It was estimated~\cite{dEnterria:2013sgr} that at $\sqrt{s}=0.9$
TeV ($\sqrt{s}=7$ TeV) the contribution due to the gluon FF dominates
over the quark one in the region $p_T^h\lesssim 20$ GeV
($p_T^h\lesssim 100$ GeV).
Therefore, the gluon contribution remains sizeable in most of the kinematic
region covered by the $pp$ measurements considered in this analysis.
For this reason we expect that including $pp$ data in a fit will have a
significant impact on the gluon FF\@.

Perturbative corrections to the hard cross sections $K_{ij}^l$ in
Eq.~\eqref{eq:ppxsec} are currently known up to
$\mathcal{O}(\alpha_s^3)$~\cite{Aversa:1988vb,
  Aurenche:1999nz,deFlorian:2002az,Jager:2002xm}, {\it i.e.} NLO\@.
Theoretical predictions are computed at this order, consistently with
those for SIA data.
The numerical computation of Eq.~\eqref{eq:ppxsec} is performed with
the code presented in Refs.~\cite{deFlorian:2002az,Jager:2002xm}.
Results have been benchmarked against the alternative {\tt INCNLO}
code~\cite{Aversa:1988vb,INCNLO:web} to a relative precision
well below the experimental uncertainties.
Parton distributions are taken as an external input from the NLO
NNPDF3.1 determination~\cite{Ball:2017nwa}.
We do not include PDF uncertainties as it has been previously
shown~\cite{dEnterria:2013sgr} that they are negligible in comparison
to FF uncertainties.

At relatively small values of $p_T^h$ ($p_T^h\lesssim 5-10$ GeV), NLO
theoretical predictions for the cross section in Eq.~\eqref{eq:ppxsec}
are affected by large uncertainties due to missing higher-order
corrections~\cite{dEnterria:2013sgr}.
A kinematic cut $\ptcut$ is therefore imposed to remove all the data
with $p_T^h<\ptcut$.
In this analysis, we choose $\ptcut=7$ GeV as a nominal cut.
This value is determined by studying the stability of the FFs and the
quality of the fit upon variations of the value of $\ptcut$ in the
range $5 \text{ GeV} \leq \ptcut \leq 10$~GeV and by varying the scale
$\mu$ by a factor of two up and down with respect to our central
choice, $\mu=p_T^h$, see Sect.~\ref{sec:pertacc}.

\section{Results}\label{sec:results}

In this section we present the results of our analysis.
First, we describe how the experimental and theoretical inputs
described in Sect.~\ref{sec:input} are combined to construct our set
of FFs, dubbed NNFF1.1h.
We present the fit quality and compare the input data set to the
corresponding theoretical predictions, focusing on the impact of
hadron\hyp{}collider measurements.
Then, we motivate our choice of the value of $\ptcut$ by investigating
the stability of the fit upon variations of $\ptcut$ and of the scale
$\mu$ used to compute the hadron\hyp{}collider cross sections.
Finally, we study the consistency of the NNFF1.1h set with the NNFF1.0
sets for identified pion, kaon and proton/antiproton FFs.

\subsection{The NNFF1.1h set}\label{sec:impact}

In this analysis, we determine the FFs of unidentified charged hadrons
in two steps.
In the first step, we construct a set of $N_{\rm rep}=2000$ equally probable
Monte Carlo FF replicas from a fit to the SIA data presented in
Sect.~\ref{sec:dataset}.
In the second step, we use this set as a prior to include the $pp$ data
presented in Sect.~\ref{sec:dataset} by means of Bayesian
reweighting~\cite{Ball:2010gb,Ball:2011gg,zahari_kassabov_2019_2571601}.
The reweighted set is then unweighted to produce an ensemble of $N_{\rm rep}=100$
equally probable Monte Carlo FF replicas.
This set forms our final deliverable result, NNFF1.1h.

The initial fit to SIA data closely follows the NNFF1.0 analysis, the
methodological details of which are extensively discussed in Sects.~4.1 and 4.3
of Ref.~\cite{Bertone:2017tyb}.
The results of this fit, which we here call NNFF1.0h, were presented in
Ref.~\cite{Nocera:2017gbk}.
The NNFF1.0h set provides a good description of its dataset, with a
total $\chi^2$ per data point of $\chi_{\rm in}^2/N_{\rm dat}=0.83$ for
$N_{\rm dat}=471$ data points (note that henceforth we will use the subscript
``$_{\rm in}$'' whenever a $\chi^2$ is computed with NNFF1.0h).
The values for the individual SIA experiments included in NNFF1.0h can
be found in Table~1 of Ref.~\cite{Nocera:2017gbk}.
A data/theory comparison is reported in Fig.~1 of the same reference.

The NNFF1.0h set is then used to produce the theoretical
predictions for the $pp$ data discussed in Sect.~\ref{sec:dataset}
according to the details presented in Sect.~\ref{sec:theory}.
The resulting values of $\chi^2_{\rm in}/N_{\rm dat}$ for each
experiment are reported in Table~\ref{tab:results}.
The corresponding data/theory comparison is displayed in
Figs.~\ref{fig:CDF}-\ref{fig:ALICE}.
The $\chi^2$ values in Table~\ref{tab:results} are computed using
the full covariance matrix, constructed from all the uncorrelated and
correlated experimental uncertainties.
For illustration the uncertainty bars shown in
Figs.~\ref{fig:CDF}-\ref{fig:ALICE} are the sum in quadrature of only the
uncorrelated uncertainties.
The effect of the correlated systematic uncertainties is taken into account
(assuming a Gaussian distribution) by shifting the theoretical
predictions~\cite{Pumplin:2002vw}.
While this shift facilitates a qualitative assessment of the data/theory
agreement, the quality of the fit can only be precisely judged from the
$\chi^2$ values reported in Table~\ref{tab:results}.

%-------------------------------------------------------------------------------
\begin{table*}[!t]
\centering
\caption{The data set included in the NNFF1.1h analysis.
  For each hadron collider experiment, we indicate the
  publication reference, the centre-of-mass energy $\sqrt{s}$, the number
  of data points included after (before) kinematic cuts $N_{\rm dat}$,
  the $\chi^2$ per number of data points before (after) reweighting,
  $\chi^2_{\rm in}/N_{\rm dat}$ ($\chi^2_{\rm rw}/N_{\rm dat}$), the number of
  effective replicas after reweighting, $N_{\rm eff}$, and the modal
  value of the $\mathcal{P}(\alpha)$ distribution in the range
  $\alpha\in [0.5,4]$, ${\rm argmax}\,\mathcal{P}(\alpha)$. For SIA
  experiments, see Table~1 in~\cite{Nocera:2017gbk}.}
\label{tab:results}
\renewcommand*{\arraystretch}{1.3}
\begin{tabularx}{\textwidth}{XXccccccc}
  \hline\noalign{\smallskip}
  Process
  & Experiment
  & Ref.
  & $\sqrt{s}$ [TeV]
  & $N_{\rm dat}$
  & $\chi^2_{\rm in}/N_{\rm dat}$
  & $\chi^2_{\rm rw}/N_{\rm dat}$
  & $N_{\rm eff}$
  & ${\rm argmax}\,\mathcal{P}(\alpha)$\\
  \noalign{\smallskip}\hline\noalign{\smallskip}
  SIA
  & \multicolumn{3}{c}{various, see Table~1 in~\cite{Nocera:2017gbk}}
  &     471   \, (527) & 0.83 & 0.83 &  ---   & --- \\
  $pp$
  & CDF   & \cite{Abe:1988yu}
  & 1.80 & \, \, 2 \, \, (49) & 3.32 & 0.20 &   1420 & 0.49 \\
  &       & \cite{Aaltonen:2009ne}
  & 1.96 & \,   50 \,   (230) & 2.93 & 1.23 & \, 735 & 1.16 \\
  & CMS   & \cite{Chatrchyan:2011av}
  & 0.90 & \, \, 7 \, \, (20) & 4.20 & 0.70 &   1206 & 0.96 \\
  &       & \cite{CMS:2012aa}
  & 2.76 & \, \, 9 \, \, (22) & 10.6 & 1.24 & \, 579 & 0.94 \\
  &       & \cite{Chatrchyan:2011av}
  & 7.00 & \,   14 \, \, (27) & 12.4 & 1.64 & \, 396 & 0.81 \\
  & ALICE & \cite{Abelev:2013ala}
  & 0.90 & \,   11 \, \, (54) & 4.94 & 1.88 &   1012 & 0.93 \\
  &       & \cite{Abelev:2013ala}
  & 2.76 & \,   27 \, \, (60) & 13.3 & 0.82 & \, 574 & 0.69 \\
  &       & \cite{Abelev:2013ala}
  & 7.00 & \,   22 \, \, (65) & 6.03 & 0.53 & \, 779 & 0.81 \\
  \noalign{\smallskip}\hline\noalign{\smallskip}
  &       & & & 603    (1054) & 6.54 & 1.11 & \, 407 & 1.10 \\
  \noalign{\smallskip}\hline
\end{tabularx}
\end{table*}
%-------------------------------------------------------------------------------

As is apparent from Table~\ref{tab:results}, the agreement between the $pp$
data and the theoretical predictions obtained with the NNFF1.0h set is not
particularly good.
The values of $\chi^2_{\rm in}/N_{\rm dat}$ range from around 3 for
the CDF data up to 13.3 for the ALICE data at $\sqrt{s}=2.76$ TeV.
However, from Figs.~\ref{fig:CDF}-\ref{fig:ALICE} we see that
theoretical predictions are affected by uncertainties due to FFs (not
included in the $\chi^2$ computation) much larger than the uncertainty
of the data.
If FF uncertainties are taken into account, the calculations based on
NNFF1.0h agree with the data at the one-$\sigma$ level.
This suggests that the $pp$ data is consistent with the SIA data used
to determine NNFF1.0h and that, at the same time, it should be able to
significantly constrain unidentified charged\hyp{}hadron FFs.

\begin{figure}[!t]
\centering
\resizebox{0.47\textwidth}{!}{
\rotatebox{270}{\includegraphics{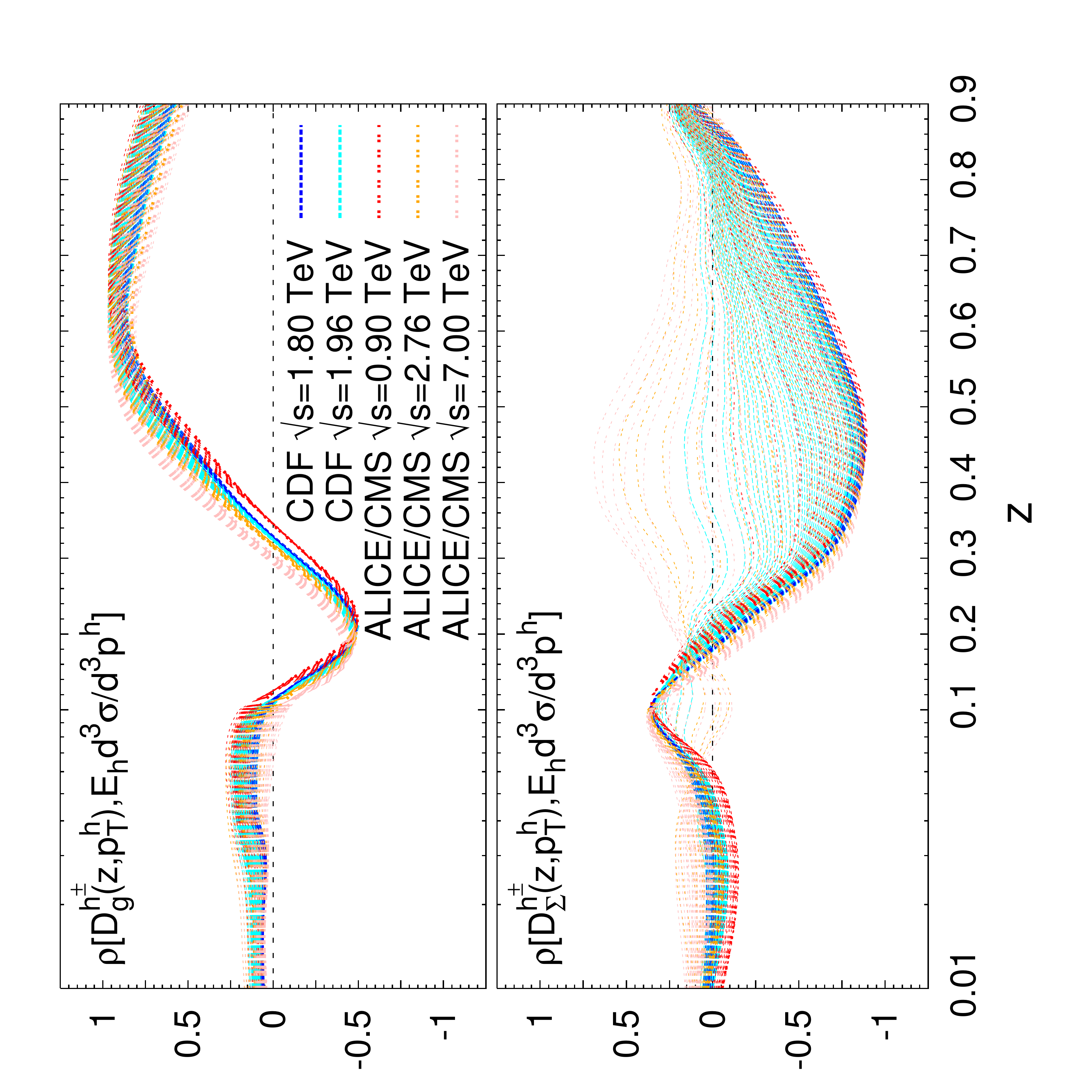}}}
\caption{The correlation coefficient $\rho$ between the gluon (top)
  and the singlet (bottom) FFs from NNFF1.0h and the
  $pp$ data listed in Table~\ref{tab:results}. Each
  data point corresponds to a separate curve; FFs are evaluated at
  a scale $\mu$ equal to the $p_T^h$ of that point.}
\label{fig:correlations}
\end{figure}

The region of the momentum fraction $z$ for which the
hadron\hyp{}collider data has potentially the largest impact on the
FFs can be quantified by computing the correlation coefficient $\rho$
(see Eq.~(1) in Ref.~\cite{Guffanti:2010yu} for its definition)
between the FFs in the NNFF1.0h set and the theoretical predictions
corresponding to the $pp$ data sets discussed in
Sect.~\ref{sec:dataset}.
Large values of $|\rho|$ indicate regions in $z$ where the sensitivity
of FFs to the data is most significant.
The correlation coefficient $\rho$ is displayed in
Fig.~\ref{fig:correlations} for the gluon and singlet FFs.
Each curve corresponds to a different data point; FFs are evaluated at
the scale $\mu$ equal to the $p_T^h$ of that point.
We observe that the correlation is maximal for $z\gtrsim 0.4$ in the
case of the gluon FF for almost all data points and for
$0.2\lesssim z\lesssim 0.7$ in the case of the singlet FF, although
for a more limited number of data points.
The sensitivity is negligible for $z\lesssim 0.1$ in both cases.

%-------------------------------------------------------------------------------
%-------------------------------------------------------------------------------

The $pp$ data listed in Table~\ref{tab:results} is used to constrain
the NNFF1.0h set by means of Bayesian
reweighting~\cite{Ball:2010gb,Ball:2011gg,zahari_kassabov_2019_2571601}.
This method consists in updating the representation of the probability
density in the space of FFs by means of Bayes' theorem.
Specifically, each replica of the NNFF1.0h set is assigned a weight
that quantifies its agreement with the new data.
These weights are computed by evaluating the $\chi^2$ of the new data
using the predictions obtained with that given replica.
After reweighting, replicas with smaller weights become less relevant in
ensemble averages, therefore the number of effective replicas in the
Monte Carlo ensemble is reduced.
The consistency of the data used for reweighting with the prior can be
assessed by examining the $\mathcal{P}(\alpha)$ profile of the new
data, where $\alpha$ is the factor by which the uncertainty of the new
data must be rescaled in order for both the prior and the reweighted
sets to be consistent with each other.
If the modal value of $\alpha$ is close to unity, the new data is
consistent with the original one within the quoted experimental uncertainties.

We construct the NNFF1.1h set by reweighting the NNFF1.0h set simultaneously
with all the $pp$ data listed in Table~\ref{tab:results}.
The values of the $\chi^2$ per data point after reweighting,
$\chi^2_{\rm rw}/N_{\rm dat}$, the number of effective replicas,
$N_{\rm eff}$, and the modal value of the $\mathcal{P}(\alpha)$
distribution in the region $\alpha\in [0.5,4]$,
${\rm argmax}\,\mathcal{P}(\alpha)$, are also collected in
Table~\ref{tab:results}.

The value of the $\chi^2$ per data point for the $pp$ data decreases
significantly after reweighting for all experiments down to values of order one.
The improvement is particularly marked for the CMS and ALICE data,
where experimental uncertainties are smaller than those for CDF\@.
The description of the SIA data is not affected by the inclusion of the $pp$
data in the fit, since the corresponding $\chi^2$ remains unchanged.
We explicitly checked that this is true also for the individual SIA experiments.
This confirms that there is no tension between the new $pp$
measurements and the SIA data used in NNFF1.0h.

The number of effective replicas after reweighting depends significantly
on the specific data set: in general, the more precise the data set,
the smaller the number of effective replicas.
The total size of the reweighted FF set, made of $N_{\rm eff}=407$
effective replicas, is around 20\% of the size of the prior set, composed of
$N_{\rm rep}=2000$ replicas.
This number is sufficiently large to ensure an adequate statistical
accuracy of the unweighted FF set, since it is significantly larger
than $N_{\rm rep}=100$, the customary number of replicas of a typical
NNPDF set.
The reweighted set is then finally unweighted into $N_{\rm rep}=100$
equally probable replicas to construct the NNFF1.1h set.

The modal value of the $\mathcal{P}(\alpha)$ distribution in the
region $\alpha\in [0.5,4]$, ${\rm argmax}\,\mathcal{P}(\alpha)$, is of
order one for all $pp$ data sets.
This is a further confirmation of the consistency within the quoted
experimental uncertainties of the $pp$ and SIA data sets included in
this analysis.

\begin{figure}[!t]
\centering
\resizebox{0.47\textwidth}{!}{
\rotatebox{270}{\includegraphics{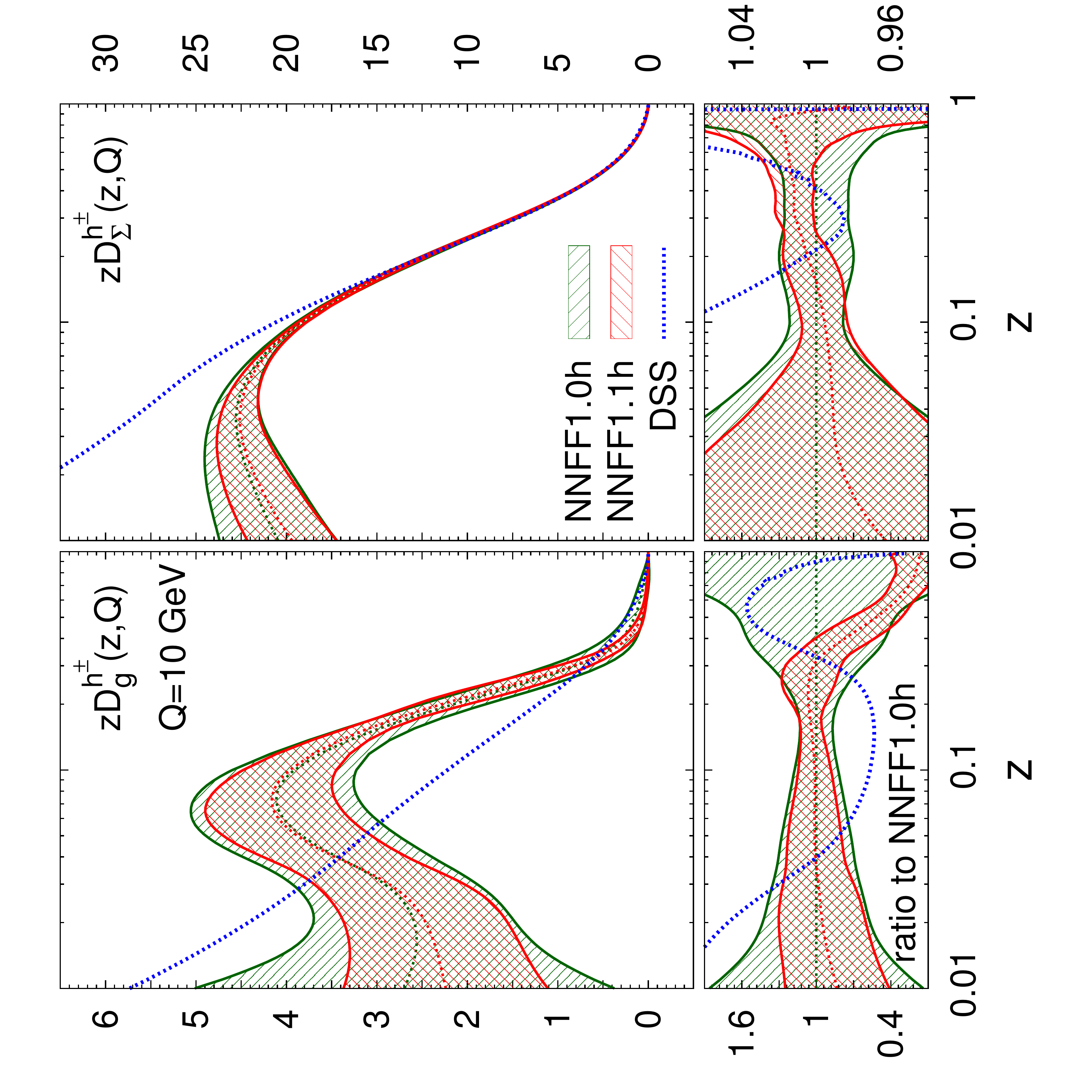}}}
\caption{The gluon (left) and singlet (right) FFs for the unidentified
  charged hadrons from NNFF1.0h, NNFF1.1h, and
  DSS at $Q=10$ GeV\@; the bands
  indicate the one-$\sigma$ uncertainties. The ratio to NNFF1.0h is
  displayed in the bottom panels.}
\label{fig:HAFFs}
\end{figure}

\begin{figure}[!t]
\centering
\resizebox{0.47\textwidth}{!}{\trimbox{0cm 9cm 0cm 1.2cm}{
\rotatebox{270}{
\includegraphics{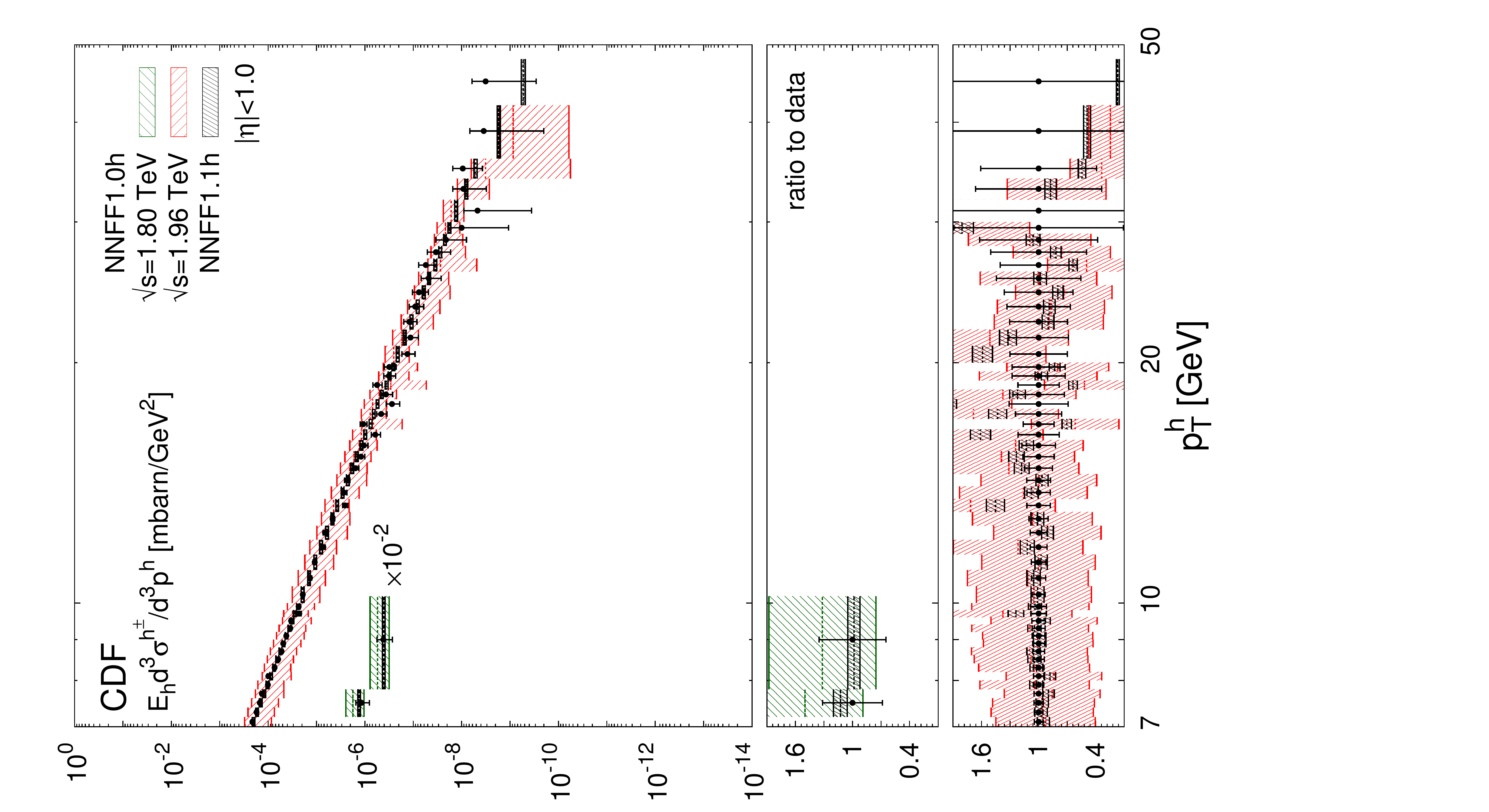}}}}
\caption{The differential cross section, Eq.~\eqref{eq:ppxsec}, for the
  inclusive charged hadron spectra measured by CDF in
  proton\hyp{}antiproton collisions at different centre\hyp{}of\hyp{}mass
  energies over the rapidity range $|\eta|<1$. The data is
  compared to the NLO predictions obtained with NNFF1.0h and NNFF1.1h.
  The corresponding theory/data ratio is shown in the lower panels.
  The bands include the one\hyp{}$\sigma$ FF uncertainty only.
  We show the sum in quadrature of the uncorrelated uncertainties on the
  data points, while correlated systematic errors are taken into account via
  shifts of the theoretical predictions (see text).}
\label{fig:CDF}
\end{figure}

\begin{figure}[!t]
\centering
\resizebox{0.47\textwidth}{!}{\trimbox{0cm 2cm 0cm 2cm}{
\rotatebox{270}{\includegraphics{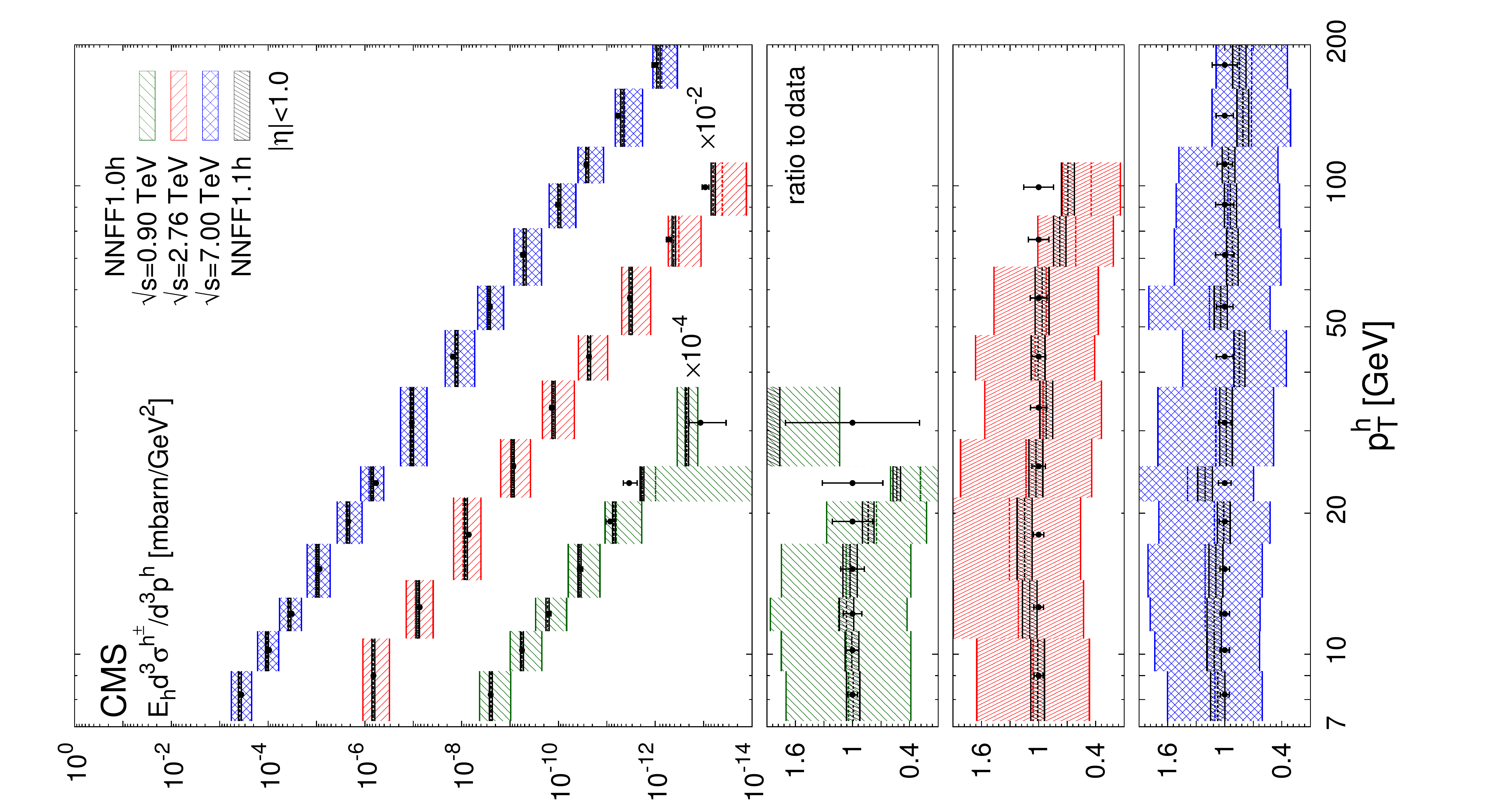}}}}
\caption{Same as Fig.~\ref{fig:CDF} for the (proton-proton) CMS data sets.}
\label{fig:CMS}
\end{figure}

\begin{figure}[!t]
\centering
\resizebox{0.47\textwidth}{!}{\trimbox{0cm 2cm 0cm 2cm}{
\rotatebox{270}{\includegraphics{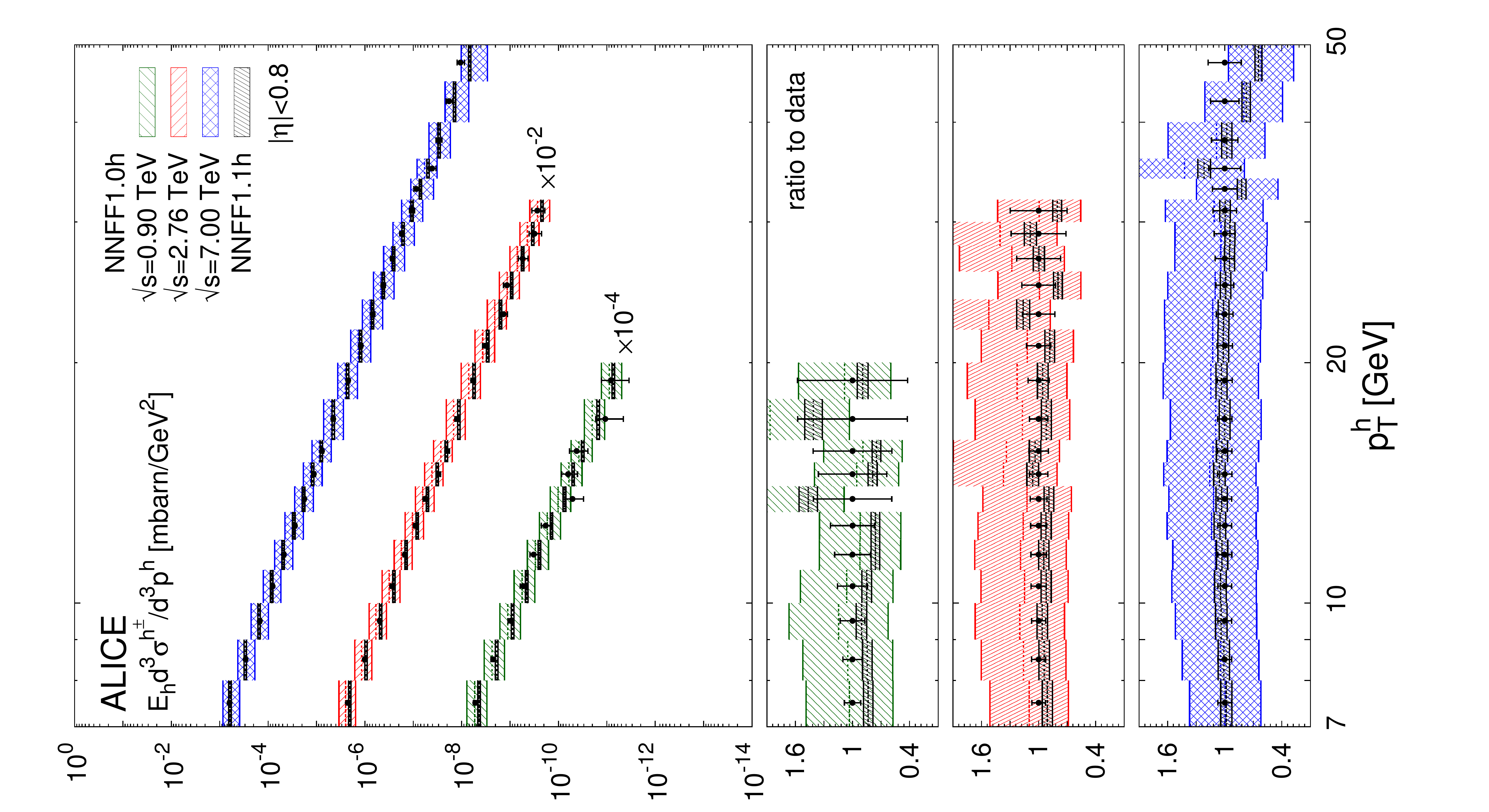}}}}
\caption{Same as Fig.~\ref{fig:CMS} for the ALICE data sets.}
\label{fig:ALICE}
\end{figure}

The gluon and singlet FFs from NNFF1.1h at $Q=10$ GeV are shown in
Fig.~\ref{fig:HAFFs}.
They are compared to the corresponding FFs from the NNFF1.0h and the
DSS~\cite{deFlorian:2007ekg} sets.
The ratio to NNFF1.0h is displayed in the bottom panel.
The theoretical predictions for the $pp$ data obtained with NNFF1.1h
are shown in Figs.~\ref{fig:CDF}-\ref{fig:ALICE} on top of their
counterparts obtained from NNFF1.0h.

%-------------------------------------------------------------------------------
%-------------------------------------------------------------------------------
%-------------------------------------------------------------------------------
%-------------------------------------------------------------------------------
%-------------------------------------------------------------------------------
%-------------------------------------------------------------------------------
%-------------------------------------------------------------------------------
%-------------------------------------------------------------------------------

As is apparent from Table~\ref{tab:results} and
Figs.~\ref{fig:HAFFs}-\ref{fig:ALICE}, the impact of the $pp$ data on the FFs
is twofold.
First, it induces a modification of the shape of the FFs.
The central value of the gluon FF moves towards slightly
harder values in the region $0.1\lesssim z \lesssim 0.3$ and towards
significantly softer values in the region
$0.3\lesssim z \lesssim 0.9$.
The central value of the singlet FF remains stable except in the
region $0.1\lesssim z \lesssim 0.4$, where it becomes slightly
smaller.
Second, the $pp$ data leads to a significant reduction of the FF
uncertainties.
For the gluon FF the relative uncertainty drops from 20\%-60\% to
10\%-15\% in the region $z\gtrsim 0.1$, {\it i.e.} a reduction of up
to a factor four.
For the singlet FF which is already well constrained by SIA data, the
reduction is more moderate but still significant, with the uncertainty
decreasing in the region $0.1\lesssim z \lesssim 0.4$ from around 2\%
to $\simeq$1\%.
Both the shape and the uncertainties of the gluon and singlet FFs are
almost unchanged for $z\lesssim 0.07$, as expected from the
correlations between $pp$ data and FFs shown in
Fig.~\ref{fig:correlations}.
The NNFF1.1h uncertainty bands are encapsulated by those of NNFF1.0h.
This further confirms the good consistency between SIA and $pp$
measurements included in our analysis.

Finally, we note that the central value of the gluon and singlet FFs
of the NNFF1.1h set is quite different from that of the DSS set.
Specifically, the gluon and singlet FFs are harder in NNFF1.1h than in
DSS for $0.03\lesssim z\lesssim 0.3$ but softer elsewhere.
No estimate of the FF uncertainties was determined in the DSS fit,
hence it is not possible to quantitatively assess its statistical
compatibility with our results.
The fact that hadron\hyp{}collider cross sections prefer a softer
gluon FF at large-$z$ was already suggested in
Ref.~\cite{dEnterria:2013sgr} as a possible explanation of the poor
agreement between $pp$ data and theory predictions when the latter is
computed with DSS\@.

\subsection{Dependence on the value of $\ptcut$}\label{sec:pertacc}

Having presented the impact of the $pp$ data on FFs, we now provide a
rationale for our choice of the baseline cut on the hadron transverse
momentum, $\ptcut = 7$~GeV.
This is motivated by examining the dependence of our study upon this cut in the
range $5$ GeV $\le \ptcut \le 10 $ GeV with steps of $1$ GeV.
This range of $\ptcut$ values being chosen in accordance with the study of
Ref.~\cite{dEnterria:2013sgr}, where it was shown that in this range
theoretical uncertainties due to missing higher\hyp{}order corrections
become sizeable.
In Table~\ref{tab:chi2s}, we collect the number of data points after the cut
and the corresponding $\chi^2_{\rm rw}/N_{\rm dat}$ values after the $pp$ data set
is used to reweight the NNFF1.0h set.

The fits with the most restrictive cuts, $\ptcut=9$~GeV and
$\ptcut=10$~GeV, naturally have a number of data points rather smaller than
those with the less conservative cut, $\ptcut=5$~GeV.
Most notably, no points of the $\sqrt{s}= 1.80$~TeV CDF data set pass
these cuts.

As one may expect the overall fit quality deteriorates, albeit modestly, if a
larger number of low-$p_T^h$ points is included in the fit.
In particular, the total $\chi^2_{\rm rw}/N_{\rm dat}$ of the $pp$ data sets
increases from $1.08$ for $\ptcut=10$~GeV to $1.27$ for $\ptcut=5$~GeV.
The description of almost all data sets is worse or significantly
worse in the fit with $\ptcut=5$~GeV than in that with
$\ptcut=10$~GeV.
For the CMS 7~TeV and ALICE 0.9~TeV data sets, the
$\chi^2_{\rm rw}/N_{\rm dat}$ increases from 1.40 and 1.52 to 2.01 and
2.56, respectively, when one lowers the cut from $10$~GeV to $5$~GeV.
The description of the ALICE 2.76~TeV and 7~TeV data is instead
moderately better with $\ptcut=5$~GeV than the one with
$\ptcut= 10$~GeV.

%-------------------------------------------------------------------------------
\begin{table*}[!t]
\centering
\begin{tikzpicture}
\label{tab:chi2s}
\small
\renewcommand*{\arraystretch}{1.3}
\node (table) {\begin{tabularx}{\textwidth}{X c cr cr cr cr cr cr}
  \hline\noalign{\smallskip}
  & $\ptcut$
  & \multicolumn{2}{c}{5 GeV}
  & \multicolumn{2}{c}{6 GeV}
  & \multicolumn{2}{c}{{7 GeV}}
  & \multicolumn{2}{c}{8 GeV}
  & \multicolumn{2}{c}{9 GeV}
  & \multicolumn{2}{c}{10 GeV}
  \\
  Experiment
  & $\sqrt{s}$ [TeV]
  & $\frac{\chi^2_{\rm rw}}{N_{\rm dat}}$ & $N_{\rm dat}$
  & $\frac{\chi^2_{\rm rw}}{N_{\rm dat}}$ & $N_{\rm dat}$
  & ${\frac{\chi^2_{\rm rw}}{N_{\rm dat}}}$ & ${N_{\rm dat}}$
  & $\frac{\chi^2_{\rm rw}}{N_{\rm dat}}$ & $N_{\rm dat}$
  & $\frac{\chi^2_{\rm rw}}{N_{\rm dat}}$ & $N_{\rm dat}$
  & $\frac{\chi^2_{\rm rw}}{N_{\rm dat}}$ & $N_{\rm dat}$\\
  \noalign{\smallskip}\hline\noalign{\smallskip}
  CDF
  & 1.80
  & 1.30 &    7
  & 0.28 &    4
  & {0.10} &    {2}
  & 0.04 &    1
  & ---  &  ---
  & ---  &  --- \\
  & 1.96
  & 1.32 &  60
  & 1.26 &  55
  & {1.23} &  {50}
  & 1.20 &  45
  & 1.15 &  40
  & 1.15 &  35 \\
  CMS
  & 0.90
  & 0.93 &  10
  & 0.67 &   8
  & {0.70} &   {7}
  & 0.71 &   7
  & 0.80 &   6
  & 0.80 &   6 \\
  & 2.76
  & 1.38 &  11
  & 1.27 &  10
  & {1.24} &   {9}
  & 1.17 &   9
  & 1.22 &   8
  & 1.16 &   8 \\
  & 7.00
  & 2.01 &  17
  & 1.80 &  15
  & {1.64} &  {14}
  & 1.52 &  14
  & 1.47 &  13
  & 1.40 &  13 \\
  ALICE
  & 0.90
  & 2.56 &  15
  & 2.05 &  13
  & {1.88} &  {11}
  & 1.71 &  10
  & 1.51 &   9
  & 1.52 &   8 \\
  & 2.76
  & 0.61 &  21
  & 0.72 &  19
  & {0.82} &  {17}
  & 0.89 &  16
  & 0.98 &  15
  & 1.08 &  14 \\
  & 7.00
  & 0.56 &  26
  & 0.52 &  24
  & {0.53} &  {22}
  & 0.55 &  21
  & 0.57 &  20
  & 0.60 &  19  \\
  \noalign{\smallskip}\hline\noalign{\smallskip}
  Total  &
  & 1.27 & 167
  & 1.14 & 148
  & {1.11} & {132}
  & 1.09 & 123
  & 1.08 & 111
  & 1.08 & 103 \\
  \hline\noalign{\smallskip}
\end{tabularx}};
\draw [black, dashed,rounded corners,thick]
  ($(table.south west) !.08! (table.north west)  !.52! (table.south east)$)
  rectangle
  ($(table.south east)  !1.52! (table.north east)  !.36! (table.south west)$);
\end{tikzpicture}
\caption{The values of the $\chi^2$ per data point, $\chi^2_{\rm rw}/N_{\rm dat}$,
  and the number of data points after cuts, $N_{\rm dat}$,
  for the $pp$ experiments included in the fit (and their total)
  for a range of values of the kinematic cut $\ptcut$.
  Our baseline is $\ptcut=7$~GeV.
}
\end{table*}
%-------------------------------------------------------------------------------

The overall fit quality turns out to be very similar for values of $\ptcut$
larger or equal to 6~GeV.
Conversely, it markedly worsens when we lower the value of $\ptcut$ from
$6$~GeV to $5$~GeV.
In this case, the $\chi^2_{\rm rw}/N_{\rm dat}$ increases from 1.14 to 1.27, mostly
because of the poor description of the 1.8 TeV CDF data set, whose
$\chi^2_{\rm rw}/N_{\rm dat}$ raises from 0.28 to 1.30.
A deterioration is also observed in the $\chi^2_{\rm rw}/N_{\rm dat}$ of almost all
the other data sets; in particular, it increases from 0.67 to 0.93
and from 2.05 to 2.56 for the 0.9 TeV CMS and ALICE data sets respectively.

This study of the fit quality suggests that reliable results require a
value of $\ptcut \geq 6$~GeV.
To find the optimal value of $\ptcut$ in the restricted range
6~GeV $\lesssim\ptcut\lesssim$ 10~GeV, we investigate the perturbative
stability of the FFs by repeating the reweighting procedure with the
scale $\mu$ in Eq.~\eqref{eq:ppxsec} set to $2 p_T^h$ and $p_T^h/2$.
We then study the behaviour of the resulting FFs for different values of
$\ptcut$.
We find that FFs are reasonably stable with respect to variations of
the scale $\mu$ if $\ptcut$ is equal to 7~GeV or larger, whereas the
same variations lead to larger distortions in shape for
$\ptcut=6$~GeV.

\begin{figure}[!t]
\centering
\resizebox{0.47\textwidth}{!}{
\rotatebox{270}{\includegraphics{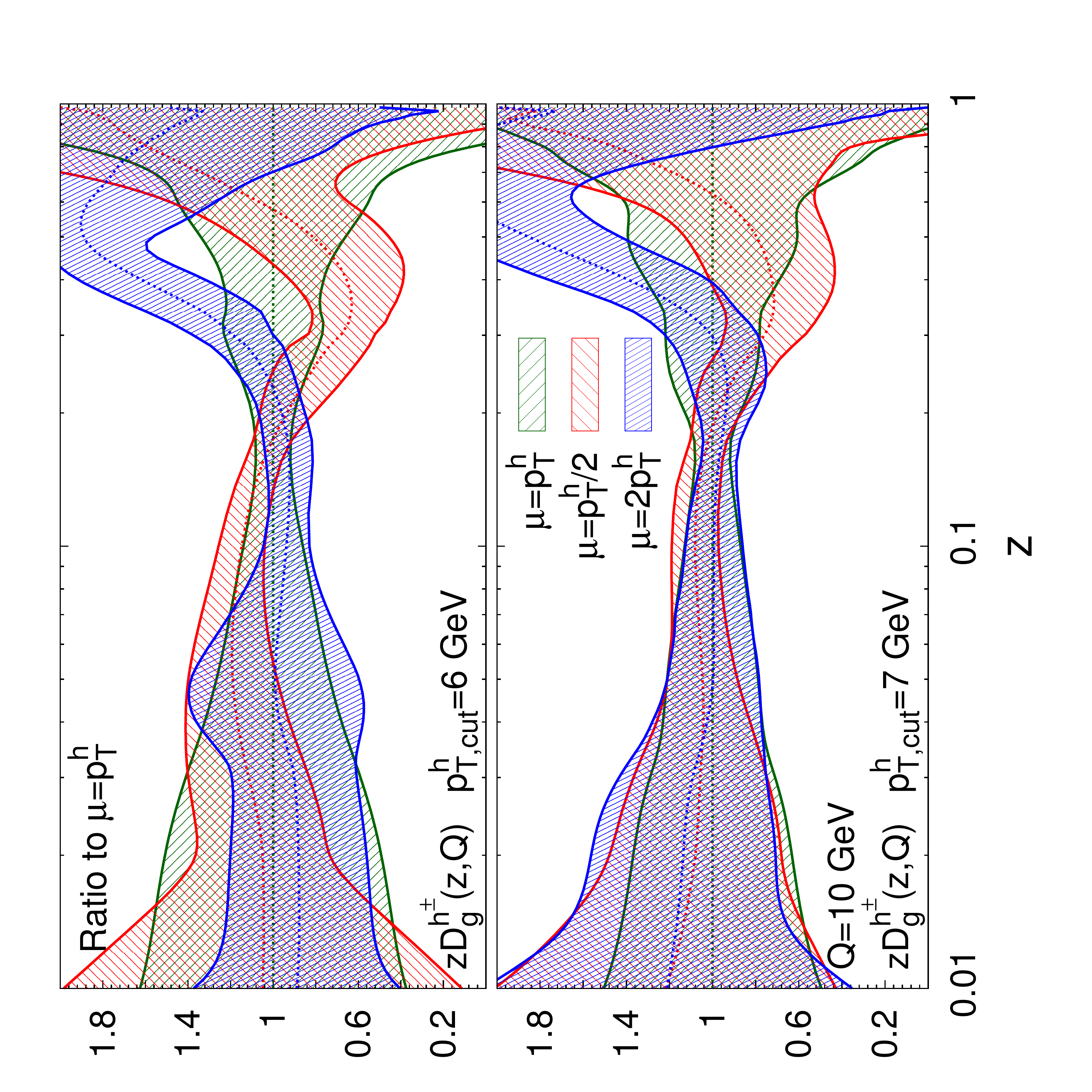}}}
\caption{Comparison of the gluon FF at $Q= 10$~GeV for the fits
  performed setting the scale $\mu$ in Eq.~\eqref{eq:ppxsec} to $p_T^h$,
  $2 p_T^h$ or $p_T^h/2$ for $\ptcut=6$~GeV (upper)
  and the baseline $\ptcut=7$~GeV (lower plot), normalised
  to the $\mu=p_T^h$ result.}
\label{fig:ratiopt7pt6sv}
\end{figure}

To illustrate this, in Fig.~\ref{fig:ratiopt7pt6sv} we show a
comparison of the gluon FF for $\ptcut=6$~GeV and $\ptcut=7$~GeV at
$Q=10$ GeV for the fits performed setting the scale $\mu$ to $p_T^h$,
$2 p_T^h$, and $p_T^h/2$, normalised to the nominal $\mu=p_T^h$
result.
We observe that in the $\ptcut=6$~GeV case, for values of $z$ between
0.1 and 0.5, the two uncertainty bands of the FFs with $\mu=2p_T^h$
and $\mu=p_T^h/2$ do not overlap, and that their central value is not
contained in the band of the FFs obtained using the central scale
$\mu=p_T^h$.
This discrepancy is partially reduced with $\ptcut=7$~GeV and we
checked that the fit with $\ptcut=10$~GeV has a similar pattern.
This behaviour is also exhibited by the singlet FF\@.
We conclude that by choosing $\ptcut=6$~GeV one would add to the fit
data points that may not be described reliably using NLO QCD theory.
Therefore, this motivates our baseline choice $\ptcut = 7$~GeV.

%-------------------------------------------------------------------------------
%-------------------------------------------------------------------------------

\begin{figure}[!t]
\centering
\resizebox{0.47\textwidth}{!}{\trimbox{0cm 0cm 0cm 0cm}{
\rotatebox{270}{\includegraphics{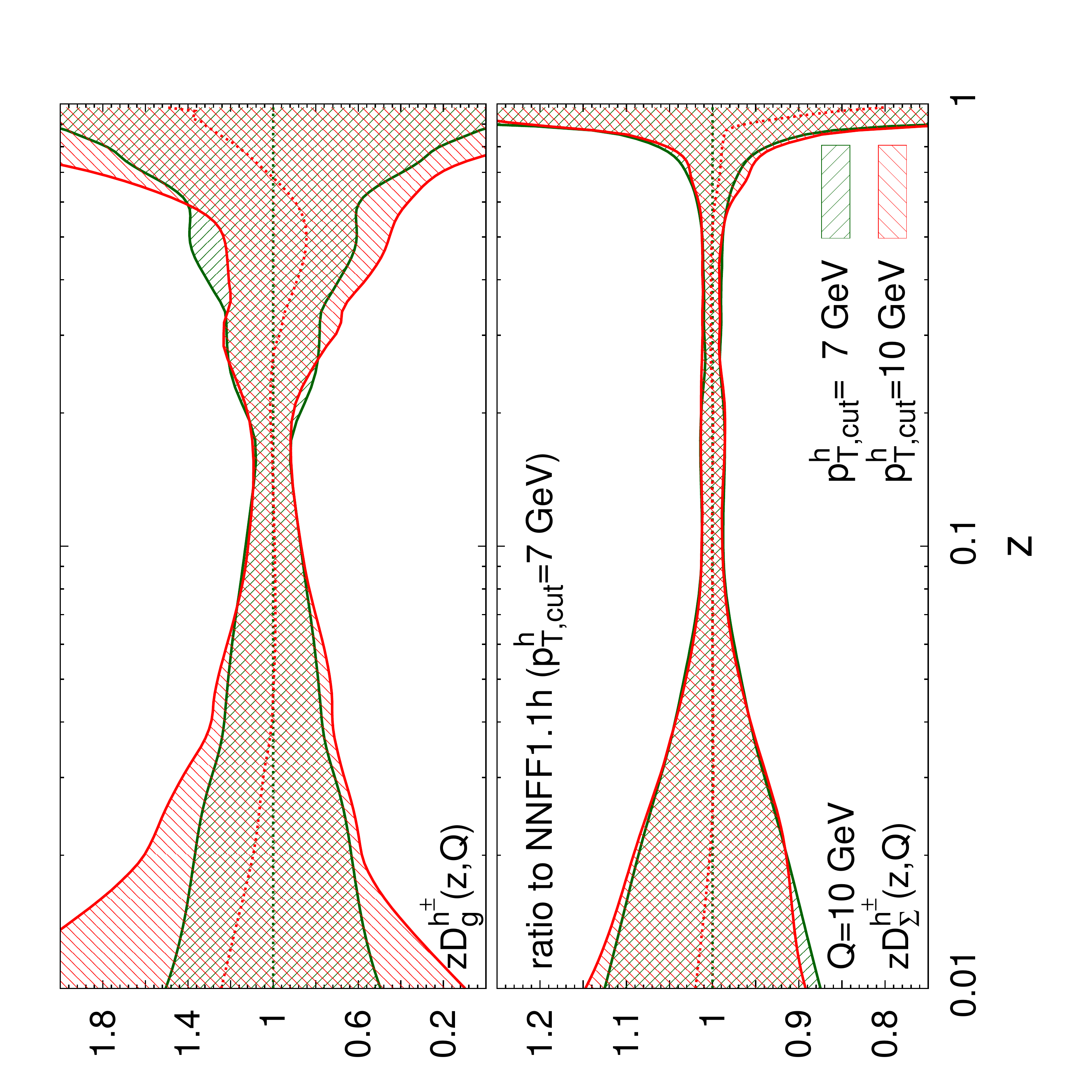}}}}
\caption{Comparison of the gluon (upper) and singlet (lower plot) FFs
at $Q=10$ GeV for the NNFF1.1h fits with $\ptcut=7$~GeV and $\ptcut=10$~GeV,
normalised to the former.}
\label{fig:ratiopt6vspt10}
\end{figure}

As further evidence in favour of our choice of $\ptcut$, in
Fig.~\ref{fig:ratiopt6vspt10} we compare the gluon and singlet FFs at
$Q=10$ GeV from the fit with our default choice $\ptcut = 7$~GeV to
those obtained with the more restrictive $\ptcut = 10$~GeV, normalised
to the former.
In both cases the resulting FFs are similar and the central value of
the $\ptcut = 7$~GeV fit is always contained within the uncertainty
band of the $\ptcut = 10$~GeV fit.
This comparison shows that the two fits are compatible and
demonstrates the reliability of the fit upon our nominal choice of
$\ptcut$.

In summary, the study of the fit quality and of the stability of FFs with
respect to scale variations suggests that the choice $\ptcut = 7$~GeV is
reasonably optimal: it allows us to include in the fit a sufficiently large
number of data points and at the same time it guarantees that the fit is not
significantly affected by missing higher\hyp{}order corrections.

%-------------------------------------------------------------------------------
%-------------------------------------------------------------------------------

\subsection{Compatibility with NNFF1.0}\label{sec:consistency}

For each parton $i$, the FFs of unidentified charged hadrons,
$D_i^{h^\pm}$, can be regarded as the sum of the FFs of charged pions,
$D_i^{\pi^\pm}$, charged kaons, $D_i^{K^\pm}$, protons and
antiprotons, $D_i^{p/\bar{p}}$, and a residual component,
$D_i^{\rm res^\pm}$, which accounts for heavier charged hadrons, such that
\begin{equation}
D_i^{h^\pm}=D_i^{\pi^\pm}+D_i^{K^\pm}+D_i^{p/\bar{p}}+D_i^{\rm res^\pm}\,.
\label{eq:hadsum}
\end{equation}
Therefore, cross sections for unidentified charged hadrons can be
expressed as the sum of individual cross sections computed with
$\pi^\pm$, $K^\pm$, $p/\bar{p}$ and residual FFs.

\begin{figure}[!t]
\centering
\resizebox{0.45\textwidth}{!}{\trimbox{0cm 0cm 0cm 0cm}{
\rotatebox{270}{\includegraphics{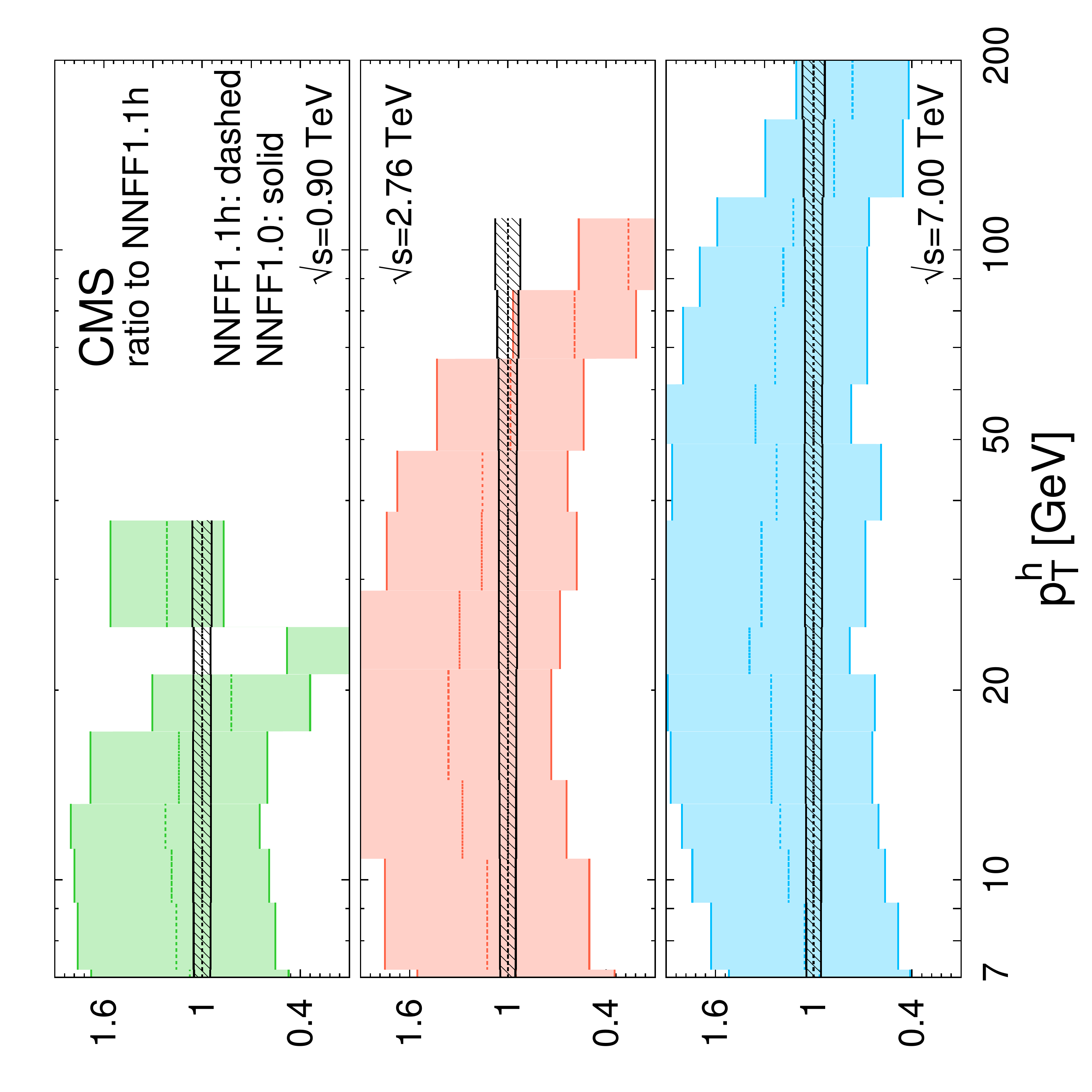}}}}
\caption{Theoretical predictions for the differential cross sections in
  $pp$ collisions, Eq.~\eqref{eq:ppxsec}, computed at NLO in the
  kinematic bins measured by CMS. We compare the predictions obtained from the
  unidentified charged hadron in the NNFF1.1h set with those obtained from
  the sum of charged pions, kaons and protons/antiprotons in the
  NNFF1.0 set. Predictions are normalised to NNFF1.1h.}
\label{fig:CMS_NN}
\end{figure}

In this work we do not use Eq.~\eqref{eq:hadsum} as a theoretical constraint
to our FF analysis, as done, for instance, in Ref.~\cite{deFlorian:2007ekg}.
The FFs for unidentified charged hadrons in NNFF1.1h are determined
independently from the FFs of identified pions, kaons and
protons/antiprotons previously obtained in NNFF1.0.
It is therefore interesting to check their consistency.
We do so by verifying that the $pp$ cross section in Eq.~\eqref{eq:ppxsec}
satisfies, within FF uncertainties, the inequality
\begin{equation}\label{eq:SIAxsecsum}
\begin{array}{rcl}
  \displaystyle E_h\frac{d^3\sigma^{h^\pm}}{d^3p^h}
  &>&\displaystyle
      \sum_{\mathcal{H}=\pi^\pm,K^\pm,p/\bar{p}}E_h\frac{d^3\sigma^{\mathcal{H}}}{d^3p^h}\,,
\end{array}
\end{equation}
which follows from the positivity of cross sections.
In Fig.~\ref{fig:CMS_NN}, we compare the l.h.s.\ and the r.h.s.\ of
Eq.~\eqref{eq:SIAxsecsum}, computed at NLO with the FFs from NNFF1.1h
and NNFF1.0, respectively, and, as a representative example, 
for the kinematics of the CMS data.
The bands in Fig.~\ref{fig:CMS_NN} correspond to one-$\sigma$ FF
uncertainties.
We assume that FFs for individual hadronic species are uncorrelated,
therefore the uncertainties for the r.h.s.\ of
Eq.~\eqref{eq:SIAxsecsum} are determined by adding in quadrature the
uncertainties from the pion, kaon and proton/antiproton NNFF1.0 sets.

The comparison in Fig.~\ref{fig:CMS_NN} shows that the inequality in
Eq.~\eqref{eq:SIAxsecsum} is always satisfied within the large
uncertainties of the NNFF1.0 result.
This also suggests that FF uncertainties for individual hadronic
species can be significantly reduced if the corresponding $pp$ data
are used in their determination.

%-------------------------------------------------------------------------------
%-------------------------------------------------------------------------------

The consistency between NNFF1.1h and NNFF1.0 can be further assessed in a
complementary way by computing the momentum carried by all charged hadrons
produced in the fragmentation of the parton (or combination of partons) $i$
and by comparing it to the same quantity computed using pions,
kaons and protons/antiprotons only.
The following relation should then hold
within uncertainties:
\begin{equation}\label{eq:moments}
\begin{array}{l}
  M_i^{h^\pm}(Q)\equiv\displaystyle\int_{z_{\rm min}}^1dz\, z
  D_i^{h^\pm}(z,Q) \gtrsim\\
  \displaystyle M_i^{\rm light}(Q)
  \equiv
  \sum_{\mathcal{H}=\pi^\pm,K^\pm,p/\bar{p}}\int_{z_{\rm min}}^1dz\, z D_i^{\mathcal{H}}(z,Q)\,.
\end{array}
\end{equation}
According to the same argument given around Eq.~\eqref{eq:SIAxsecsum},
the momentum carried by heavier charged hadrons has to be positive.
However, contrary to Eq.~\eqref{eq:SIAxsecsum}, the inequality does not have
to be strictly fulfilled as the integration over $z$ in
Eq.~\eqref{eq:moments} is truncated at $z_{\rm min}$ due to the impossibility
of determining FFs down to very small values of $z$.
Therefore these (truncated) momentum fractions are not guaranteed to be
strictly positive.

We compute $M_i^{h^\pm}(Q)$ and $M_i^{\rm light}(Q)$ in Eq.~\eqref{eq:moments}
using $z_{\rm min}=0.01$ and $Q=5$ GeV for NNFF1.1h and NNFF1.0 for charged
pions, charged kaons, and protons/antiprotons.
The uncertainty of $M_i^{\rm light}(Q)$ is determined by adding in quadrature
the uncertainties obtained from the single NNFF1.0 sets.
The resulting momentum fraction of the gluon FF and the $u^+$, $d^++s^+$,
$c^+$ and $b^+$ combinations of quark FFs, with $q^+\equiv q+\overline{q}$,
are reported in
Table~\ref{tab:moments}.
For all the parton combinations considered, $M_i^{h^\pm}(Q)$ and
$M_i^{\rm light}(Q)$ are compatible within the FF errors, hence the inequality
in Eq.~\eqref{eq:moments} is not violated.
We therefore conclude that the NNFF1.1h and NNFF1.0 sets are consistent.

We note that the uncertainties of the truncated moments computed with NNFF1.1h 
are about a factor of three smaller than those obtained with NNFF1.0.
This reduction highlights once more the significant constraining power
of the $pp$ data on the FFs.
Additionally, the central value of $M_i^{h^\pm}$ is in general only slightly
larger than that of $M_i^{\rm light}$ (except for $u^+$ and $d^++s^+$).
This suggests that the momentum fraction carried by charged hadrons
other than pions, kaons and protons/anti\hyp{}protons is small and
within the uncertainties of NNFF1.1h.

%-------------------------------------------------------------------------------
\begin{table}[!t]
\centering
\caption{The momentum fraction, Eq.~\eqref{eq:moments}, for the gluon,
  $u^+$, $d^++s^+$, $c^+$ and $b^+$ FF combinations computed at $Q=5$ GeV and
  $z_{\rm min}=0.01$ for the unidentified charged hadron FFs from NNFF1.1h and
  for the sum of charged pion, kaon and proton/antiproton FFs from NNFF1.0.}
\label{tab:moments}
\renewcommand*{\arraystretch}{1.3}
\begin{tabularx}{\columnwidth}{XUU}
  \hline\noalign{\smallskip}
  $Q=5\text{ GeV}$
  & NNFF1.1h
  & NNFF1.0  \\
  $i$ & $M_i^{h^\pm}(Q)$
  & $M_i^{\rm light}(Q)$ \\
  \noalign{\smallskip}\hline\noalign{\smallskip}
  $g$
  & $0.86\pm 0.06$ & $0.80\pm 0.18$\\
  $u^+$
  & $1.24\pm 0.07$ & $1.42\pm 0.12$\\
  $d^++s^+$
  & $2.05\pm 0.08$ & $2.07\pm 0.27$\\
  $c^+$
  & $1.09\pm 0.03$ & $1.01\pm 0.08$\\
  $b^+$
  & $1.06\pm 0.02$ & $0.98\pm 0.08$\\
  \hline\noalign{\smallskip}
\end{tabularx}
\end{table}
%-------------------------------------------------------------------------------

\section{Summary and outlook}\label{sec:summary}

In this work we presented NNFF1.1h, a new determination of the FFs of 
unidentified charged hadrons based on a comprehensive set of SIA and $pp$ 
measurements.
Our study demonstrates that all the data can be simultaneously very
well described and that $pp$ data significantly constrains the so far
poorly known gluon FF.
The robustness of NNFF1.1h against potentially large missing
higher-order perturbative corrections in the $pp$ predictions was
ensured by appropriate kinematic cuts.
Specifically, the reliability of our results upon our choice of the
kinematic cut on the hadron transverse momentum was explicitly verified.
We also demonstrated that the NNFF1.1h set is consistent with our
previous NNFF1.0 sets for identified charged pions, kaons and
protons/antiprotons.
Given the high precision of its gluon FF, the NNFF1.1h set could be
used to compute theoretical predictions for single\hyp{}inclusive
hadron production in proton\hyp{}ion and ion\hyp{}ion collisions,
where gluon fragmentation also dominates.

Our work could be extended in at least three directions.
First, the charged hadrons SIDIS multiplicities measured by the
COMPASS Collaboration~\cite{Adolph:2013stb,Aghasyan:2017ctw} could be
included in our analysis of unidentified charged-hadron FFs in order
to achieve flavour separation. This is possible thanks to the
sensitivity of the SIDIS observable to different FF combinations as
compared to SIA and $pp$.

Second, this analysis could be repeated for the identified hadronic
species determined in NNFF1.0.
This would be particularly well motivated in view of the increasing
amount of precise data becoming available from LHC
experiments~\cite{Abelev:2014laa,Sirunyan:2017zmn,Acharya:2017tlv}.
These measurements will complement the existing data from
RHIC~\cite{Adare:2007dg,Adams:2006nd,Abelev:2009pb,Agakishiev:2011dc,
  Adamczyk:2013yvv}, part of which, however, comes from longitudinally
polarised $pp$ collisions.
Including data on charged pion, kaon and proton production from the
LHC should lead to an improved determination of their gluon FF in the
large-$z$ region, as is the case for unidentified charged hadrons.

Finally, possible future work is motivated by the realisation that, as
shown in this analysis, the LHC data significantly improves the
precision with which FFs can be determined.
At this point, theoretical uncertainties on hadron\hyp{}collider cross
sections, such as those from missing higher orders, can become
comparable in size to the experimental uncertainties.
The calculation of NNLO QCD corrections to the $pp$ cross sections
will therefore be of increasing importance.
While such calculations are currently unavailable, they may emerge
through the work recently carried out for jet
production~\cite{Ridder:2013mf,Currie:2013dwa,
  Currie:2016bfm,deFlorian:2013qia}.
Meanwhile, our analysis could be extended by taking into account other
sources of uncertainty, such as PDF uncertainties, 
following the procedure outlined in Ref.~\cite{Ball:2018odr}.\vspace{30pt}

The NNFF1.1h set presented in this work is available through the {\tt
  LHAPDF6} interface~\cite{Buckley:2014ana}, where the required
flavour separation is generated according to the procedure for kaons
described in Appendix~A of Ref.~\cite{Bertone:2017tyb}.

%Acknowledgements
\section*{Acknowledgements}

V.B., N.P.H, J.R and L.R. are supported by the European Research Council 
Starting Grant PDF4BSM. 
J. R. is also supported by the Netherlands Organisation 
for Scientific Research (NWO).
E.R.N. is supported by UK STFC grants ST/L000458/1 and ST/P0000630/1.

%Bibliography
\bibliographystyle{spphys}
\bibliography{nnff11h}

\end{document}